\documentclass[12pt]{iopart}
\usepackage{amssymb}
\usepackage{graphicx}

\begin{document}
\title{Hydrodynamic approach to transport and
turbulence in nanoscale conductors}
\author{R. D'Agosta and M. Di Ventra}
\address{Department of Physics, University of California - San
Diego, San Diego, CA 92093}
\date{\today}

\begin{abstract} The description of electron-electron interactions in transport problems is both
analytically and numerically difficult. Here we show that a much simpler description of electron transport in
the presence of interactions can
be achieved in nanoscale systems. In particular, we show that the electron flow in nanoscale
conductors can be described by Navier-Stokes type of equations with an effective electron viscosity, i.e.,
on a par with the dynamics of a viscous and compressible classical
fluid. By using this hydrodynamic approach we derive the conditions
for the transition from laminar to turbulent flow in nanoscale systems
and discuss possible experimental tests of our predictions.
\end{abstract}

\section{Introduction}
The electron liquid is both viscous and compressible; properties
which suggest an intriguing analogy with a classical
liquid~\cite{Landau6}. This analogy is even more compelling when one
recalls that the time-dependent many-body Schr\"odinger equation
(TDSE) can be cast, quite generally, in a ``hydrodynamic'' form in
terms of the density and expectation value of the velocity
operator~\cite{Martin1959,Tokatly2005a}. In the classical case one
can derive time-dependent equations, called Navier-Stokes equations,
for the velocity field of the fluid as a function of its density,
visco-elastic coefficients, pressure and the geometric
confinement~\cite{Landau6}. These equations are centrefold in
hydrodynamics and describe both laminar and turbulent regimes. If we
could derive similar equations in the quantum case we would have a
powerful tool to investigate a plethora of effects related to
electron-electron interactions on a much simpler level than solving
for the many-body Schr\"odinger equation. Unfortunately, the
derivation of these equations in the quantum case is generally not
possible.

In this paper we show that transport in nanoscale conductors
satisfies the conditions to derive quantum Navier-Stokes
equations. This is simply due to the geometric constriction
experienced by electrons flowing in a nanostructure which gives rise
to very fast ``collisional'' processes~\cite{DiVentra2004a,
Bushong2005}. In this regime, we show that one can truncate the
infinite hierarchy of equations of motion for the electron stress
tensor to second order and thus derive quantum hydrodynamic
equations. With these equations we first rederive conductance
quantization in a quasi-1D non-viscous, incompressible fluid thus
making the connection between quantum transport and this hydrodynamic
picture clearer. We then predict the conditions for the
transition from laminar to turbulent flow in quantum point contacts
(QPCs) and suggest specific experiments to verify our predictions.

Let us start from a general many-body Hamiltonian
$\hat{H}=\hat{T}+\hat{W}+\hat{V}_{ext}$, where $\hat{T}$ is the
kinetic term, $\hat{V}_{ext}$ an external potential and
\begin{equation}
\hat{W}=\frac{1}{2}\int dr \int dr'
\psi^{\dag}(r)\psi^{\dag}(r')w(|r-r'|)\psi(r')\psi(r),
\end{equation}
where $\psi(r)$ are field operators and $w(|r-r'|)$ is the Coulomb
interaction potential. It is well known that the TDSE can be
equivalently written as two coupled equations of motion for the
single-particle density, $n(r,t)$, and velocity field, $v(r,t)$, as
obtained from the Heisenberg equation of motion for the
corresponding operators~\cite{Martin1959,Tokatly2005a}. The single
particle density is defined in terms of the field operator as
$n(r,t)=\langle \psi^\dagger(r,t)\psi(r,t)\rangle$, while the
velocity is given by $v(r,t)=j(r,t)/n(r,t)$ where the current
density is given by $j(r,t)=e\hbar\langle[
\psi^\dagger(r,t)\nabla\psi(r,t)- \nabla\psi^\dagger(r,t)
\psi(r,t)]\rangle/2mi$. For clarity we rewrite the Heisenberg
equations of motion here (summation over repeated indexes is
understood)
\begin{eqnarray}
&&D_{t}n(r,t)+n(r,t)\nabla v(r,t)=0\\
&&m n(r,t)
D_{t}v_{j}(r,t)=-\nabla_{i}P_{i,j}(r,t)-n(r,t) \nabla_{j}V_{ext}(r,t)
\label{equationmotion}
\end{eqnarray}
where $D_{t}=\partial_{t}+v\cdot \nabla$ is the
convective derivative, $m$ the electron mass, and $P_{i,j}$
($i(j)\equiv r_{i(j)} =x,y,z$) is a stress tensor, exactly given by
the sum of kinetic and interaction stress
tensors~\cite{Martin1959,Tokatly2005a}. The interaction tensor is
\begin{equation}
\label{interation}
W_{i,j}(r,t)=-\frac12\int
dr'~\frac{r'_{i}r'_{j}}{|r'|}\frac{\partial w(|r'|)}{\partial |r'|}\int_{0}^{1}d\lambda~G_{2}(r+\lambda r',r-(1-\lambda)r')
\end{equation}
where $G_{2}(r,r')=\langle \psi^{\dagger}(r)\hat n(r')
\psi(r)\rangle$ is the two-particle density matrix and $\lambda$ a
parameter that defines the geodesic which connects two interacting
particles~\cite{Tokatly2005a}. As mentioned above,
(\ref{equationmotion}) has an appealing ``hydrodynamic'' form where
all many-body information is included in the stress tensor
$P_{i,j}$. In order to solve (\ref{equationmotion}) one proceeds by
calculating an equation of motion for $G_{2}$, which can be derived
from the Heisenberg equation of motion of the particle creation and
destruction operators. However, this equation of motion contains the
three-particle density matrix. In turn, the equation of motion for
the three-particle density matrix contains the four-particle density
matrix and so forth, thus generating an infinite hierarchy of nested
equations, making the problem practically
unsolvable~\cite{Martin1959,Tokatly2005a}.

\section{Quantum Navier-Stokes equations} We show here that in the case
of electrical transport in nanoscale systems we can instead close
this set of equations. We proceed as follows. First, let us derive
the dependence of the stress tensor $P_{i,j}$ on the rate at which
the system reaches a quasi-steady state. We can obtain this
dependence from the quantum kinetic equations for the nonequilibrium
distribution function $f(r,p,t)$ ($p$ is the momentum), which can be
derived from the TDSE equation with standard
techniques,~\cite{Kadanoff}\footnote{Clearly, for the definition of
local equilibrium distribution to be valid any length scale entering
the problem has to be larger than the system Fermi wavelength.} in a
co-moving (Lagrangian) reference frame moving with velocity
$v(r,t)$~\cite{Tokatly1999}
\begin{eqnarray}
I[f]=&&D_{t}f(r,p,t)+\frac{p}{m}\nabla f(r,p,t)+e\nabla
\varphi \frac{\partial f(r,p,t)}{\partial p}
-p\cdot \nabla
v\frac{\partial f(r,p,t)}{\partial p}\nonumber\\
&&-mD_{t}v\frac{\partial
f(r,p,t)}{\partial p}
\label{kinetic}
\end{eqnarray}
where $I$ is the usual collisional integral~\cite{Kadanoff},
$\varphi$ is the sum of the external potential and the Hartree part of
the interaction potential. The collisional integral contains two
terms, one elastic and the other inelastic. In what follows, it is
important to realize that both terms can drive the system toward a
local equilibrium configuration.

The first two moments of the distribution give the density
$n(r,t)=\sum_{p}f(p,r,t)$ and the condition $\sum_{p}p f(p,r,t)=0$.
The two-particle stress tensor is related to the distribution function as $P_{i,j}=\sum_{p}p_{i}p_{j}f(p,r,t)/m$. Similarly, higher
order moments of the distribution function produce higher order stress tensors.
Introducing these moments into (\ref{kinetic}) and comparing with
(\ref{equationmotion}) we can write the stress tensor in terms of
the collisional integral $I$ according to the equation of
motion
\begin{equation}
\frac{1}{m}\int dp~ I[f]
p_ip_j=D_{t}P_{i,j}+P_{i,j}\nabla \cdot v+P_{i,k}\nabla _{k}v_{j}
+P_{k,j}\nabla_{k} v_{i}+\nabla_{k}P_{i,j,k}^{(3)},
\label{secondmoment}
\end{equation}
where $P^{(3)}$ is the three-particle stress tensor. By
writing the equation of motion for $P^{(3)}$ we would again get an
infinite hierarchy of equations.
It is interesting to point out that the theorems of
time-dependent density-functional theory establish that the stress
tensor $P_{i,j}$ is a universal functional of the velocity and density
only (see for example, \cite{Gross1996} or \cite{GiulianiVignale}).  This implies that the hierarchy of equations for the moments
of the distribution function can be formally closed to all orders in
the electron-electron interaction.
However, we note that
$P^{(3)}$ enters in (\ref{secondmoment}) only through its spatial
derivative.  If the latter is small then the hierarchy can be
truncated \cite{Tokatly1999}. From (\ref{secondmoment}) we easily
see that this derivative is small compared to the other terms whenever $\gamma=u/(L\max(\omega,\nu_{c})) \ll 1$. Here $u$ is the average electron
velocity, $L$ is the length of inhomogeneities of the liquid that give rise to scattering among
three particles, $\omega$ the system
proper frequency and $\nu_{c}$ the collision rate.
The parameter $1/L$ enters through the spatial
derivative of $P^{(3)}$, $\omega$ from the frequency dependence of the
interactions (in the DC limit of interest here $\omega\rightarrow 0$), $\nu_c$ through the collisional integral
$I[f]\propto -\nu_c(f-f_0)$, where $f_0$ is the equilibrium Fermi
distribution.
This derivative is indeed small for transport in nanostructures: When
electrons move into a nanojunction they adapt to the given junction
geometry at a fast rate, and produce a quasi-steady state and local
equilibrium distributions even in the absence of electron
interactions~\cite{DiVentra2004a, Bushong2005}. This ``relaxation''
mechanism occurs roughly at a rate $\nu_c= (\Delta t)^{-1}\sim (\hbar
/\Delta E)^{-1}$, where $\Delta E$ is the typical energy spacing of
lateral modes in the junction. For a nanojunction of width $w$ we have
$\Delta E\sim \pi^{2}\hbar^{2}/m w^{2}$ and $\Delta t \sim m
w^{2}/\pi^{2}\hbar$. If $w$ = 1 nm, $\nu_c$ is of the order of
$10^{15}$ Hz, i.e., orders of magnitude faster than typical
electron-electron or electron-phonon scattering rates. The condition
$\gamma=u/(L\max(\omega,\nu_{c}))\ll 1$ thus requires the length of inhomogeneities
$L\gg 1~\mathrm{nm}$, which is easily satisfied in
nanostructures. Note instead that in mesoscopic structures this
condition is not necessarily satisfied. In that case, the dominant
relaxation rate $\nu_c$ is given by inelastic effects, i.e. it is of
the order of THz, so that for typical lengths of mesoscopic systems,
$\gamma \approx 1$ in the DC limit. Nonetheless, the above condition
could still be valid for high-frequency excitations, like plasmons, and/or
very low densities, so that stress tensors of order higher than two are negligible.

Neglecting $\nabla_{k}P_{i,j,k}^{(3)}$ in (\ref{secondmoment}) we can thus derive a form for $P_{i,j}$. Let
us write quite generally the stress tensor $P_{i,j}$ as
$P_{i,j}=\delta_{i,j}P-\pi_{i,j}$, where the diagonal part gives the
pressure of the liquid, and $\pi_{i,j}$ is a traceless tensor that
describes the shear effect on the liquid. From (\ref{secondmoment}) we thus find that the tensor $\pi_{i,j}$ can
be written as (in $d$ dimensions, $d>1$)
\begin{equation}
\pi_{i,j}=\eta
\left(\nabla_{i}v_{j}+\nabla_{j}v_{i}-\frac{2}{d}\delta_{i,j}\nabla_{k}v_{k}\right)
\label{pixcstatic}
\end{equation}
where $\eta$ is a real coefficient that is a functional
of the density~\cite{Tokatly1999}.
We point out that (\ref{pixcstatic}) is in fact a
particular case of a general stress tensor with memory effects taken
into account~\cite{Conti1999,Vignale1996, Tokatly2005b}. In our derivation
this is the first non-trivial term of an expansion of the stress tensor in terms of the density and velocity field. Consequently the Navier-Stokes
stress tensor in (\ref{pixcstatic}) can be seen as the first-order
(non-trivial) contribution to the exact stress tensor of the electron
liquid (see also ~\cite{Tokatly2005a, Tokatly2005b, Vignale1996}).

Using this stress tensor we finally get from
(\ref{equationmotion}) the generalized Navier-Stokes equations for
the electron liquid in nanoscale systems
\begin{eqnarray}
&&D_{t}n(r,t)=-n(r,t)\nabla \cdot v(r,t),\nonumber\\
&&mn(r,t)D_{t}v_{i}(r,t)=-\nabla_{i}P(r,t)+\nabla_{j}\pi_{i,j}(r,t)
-n(r,t)\nabla_{i}V_{ext}(r,t).
\label{completeNS}
\end{eqnarray}
Equations~(\ref{completeNS}) are formally equivalent to their
classical counterpart~\cite{Landau6} and thus describe also
nonlinear solutions, i.e. the possibility to obtain turbulence of
the electron liquid in its normal state.  In the examples that
follow, we will consider only the case in which the liquid is
incompressible so that the viscoelastic coefficients are spatially
uniform: This approximation is practically satisfied in metallic
QPCs but needs to be relaxed in the case of QPCs with
organic/metallic interfaces (see e.g. \cite{Sai2005}). In addition,
for this case the Hartree potential is constant and its spatial
derivative is thus zero.  Therefore, (\ref{completeNS}) reduce to
the Navier-Stokes equations for the density and velocity of a
viscous but incompressible electron liquid
\begin{eqnarray}
&&D_{t} n(r,t)=0,\nonumber\\
\label{NS}
&&\nabla\cdot v(r,t)=0,\\
&&mn(r,t)D_{t}v_{i}(r,t)=-\nabla_{i} P(r,t)+\eta \nabla^{2}v_{i}(r,t)-n(r,t)\nabla_{i}V_{ext}(r,t).\nonumber
\end{eqnarray}

\section{Conductance quantization from hydrodynamics}
Let us first show
that we can derive from (\ref{NS}) the quantized conductance of
an ideal ($\eta= 0$) quasi-1D liquid. We consider the electron liquid
adiabatically connected to two reservoirs, and we call $v_{L(R)}$ and
$\mu_{L(R)}$ the velocity and chemical potential, respectively, in the
left (right) reservoir, with $\mu_L-\mu_R=eV_{bias}$. From
(\ref{NS}) we then derive the Bernoulli's equation that states
the conservation of energy
\begin{equation}
\frac{v_{L}^{2}}{2}+h_{L}+\frac{\mu_{L}}{m}=\frac{v_{R}^{2}}{2}+h_{R}+\frac{\mu_{R}}{m}
\label{bernoulli}
\end{equation}
where $h_{L(R)}$ is the enthalpy of the left (right)
leads\footnote{To derive (\ref{bernoulli}) one makes use of the
relation $v\nabla\cdot v=\nabla v^{2}/2-v\times(\nabla\times v)$ and
by projecting the equation of motion on the tangent to the current
flow. The enthalpy is defined as $h=P/n$~\cite{Landau6}. In the 1D
case $\nabla\times v\equiv 0$ and $h(n)=(\pi\hbar n)^2/2m^2$.} .
Since we assume the fluid is incompressible $h_{L}=h_R$. By defining
the flow velocity $v=(v_{R}+v_{L})/2$ and the co-moving Fermi
velocity\footnote{$v_{F}$ is the Fermi velocity in the reference
frame moving with velocity $(v_{R}+v_{L})/2$. Obviously the
Bernoulli's equation is invariant under any Galilean
transformation.} $v_{F}=(v_{L}-v_{R})/2$ we get from
(\ref{bernoulli}) the relation $2m v v_{F}=eV_{bias}$. By
definition, the current is given by $I=e n v$ so that, by using the
1D density of states, $I=e m v v_{F}/\pi\hbar=e^{2}V_{bias}/h$,
which, in the linear regime, gives the quantized conductance (per
spin) $G_0=I/V_{bias}=e^{2}/h$. If we assume that only a fraction
$T$ of electrons is transmitted due to the presence of a barrier in
the liquid, we can argue that, in linear response, the current is an
equal fraction of the current in the absence of the barrier, i.e.
$I=e n v T$. The conductance is thus $G=Te^{2}/h$ in accordance with
the Landauer two-terminal result~\cite{Landauer1989}. Corrections to
this conductance in the presence of viscosity in 3D have been
estimated in \cite{Sai2005} and were found to depend non-linearly on
the gradient of the electron density. Corrections to this conductance have been experimentally observed and justified with a hydrodynamical model where the conductance depends on the physical properties of the electron flow \cite{deJong1995}.

\section{Turbulence} We know that the time-independent solutions
of the Navier-Stokes equations~(\ref{completeNS}) or~(\ref{NS}) can
describe many different regimes, with the non-linear (turbulent)
regime generally favoured with respect to the laminar one.  In fluid
mechanics, in order to identify these regimes, it is customary to
define a key quantity, the Reynolds number $Re$, as the only
non-dimensional quantity that can be constructed out of the physical
parameters of the system, like the density, the viscosity $\eta$,
etc.~\cite{Landau6}. In the quantum case we follow a similar
convention. For instance in 2D we define $Re$ as $Re={Q}/{\eta}={m
I}/{e\eta}$ where $Q$ is the total mass current, and $I$ is the
average total electrical current. For small $Re$ the stable (and
stationary) flow is usually laminar while for large $Re$ the flow is
turbulent~\cite{Landau6}. In the latter case, one should then
observe a local velocity field which varies in space in an irregular
way, and whose pattern is very sensitive to initial conditions.

\begin{figure}[t]
\includegraphics[width=8cm,clip]{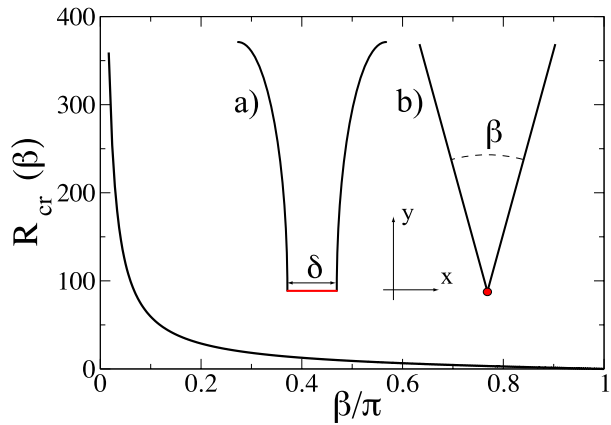}
\caption{Critical $Re$ number as a function of the angle $\beta$ of
the geometry represented in the inset (b). Inset (a) is a schematic of
an adiabatic constriction.}
\label{fig1}
\end{figure}

{\it Adiabatic QPCs -} Let us apply these concepts to the
transition between laminar and turbulent flow in QPCs. The microscopic
geometry of these structures is quite complicated so that analytical
solutions to (\ref{NS}) cannot be generally found.  However, for
an adiabatic 2D constriction, the self-consistent confining electron
potential which enters (\ref{NS}) can be approximated with the
simple form $y=k \sqrt{|x|-\delta}$ where $k >0$ is a parameter that
controls the rate at which the constriction opens up and $\delta>0$ is
the opening of the constriction [see inset (a) of Fig.~\ref{fig1}]. We
assume electrons either originate from, or enter into, the region
$\{y=0, x=[{0,\delta}]\}$ of that potential to resurface on the
opposite side of the structure. We know from classical hydrodynamics
that, for any constriction, turbulent flow may exist only on the side
where the constriction acts as a source of
electrons~\cite{Landau6}. We will therefore discuss only this side.
We are interested here in the case of $k$ large, i.e. true adiabatic
limit. Since we are dealing with a viscous liquid we assume a no-slip
condition at the boundary, i.e., at the boundary the velocity
component parallel to the boundary is zero.
This condition can be relaxed by assuming that the
velocity be finite at the boundary: The conclusions would be
unchanged. We want to point out, however, that due to the small
viscosity of the electron liquid the effective boundary conditions on
the electron liquid do not affect considerably its bulk motion.

The analytical stationary solution of (\ref{NS}) with these boundary
conditions is not known. However, one can find an approximate
solution by applying the transformation $x=p$, $y=\sqrt{k q}$ that
maps $y=k \sqrt{|x|-\delta}$ into $q=\sqrt{k} (|p|-\delta)$, $q\ge0$
and transforms (\ref{NS}) in a new set of equations where the
parameter $1/k$ appears explicitly. We can then expand this equation
in powers of $1/k$.  The zeroth order solution has the form
$v_{p}(p,q)=0$, $v_{q}(p,q)=\alpha q^{1/2}(p^{2}-\delta^{2})$ where
$v_{p(q)}$ is the $p$ $(q)$ component of the velocity in the $p-q$
plane and $\alpha$ is an arbitrary constant fixed by the requirement
that a certain amount of charge is flowing through the system. If we
transform back to the $x-y$ plane we see that the zeroth order
solution is given by $v_{x}(x,y)=0$, $v_{y}(x,y)=\alpha
(x^{2}-\delta^{2})$, i.e., we have obtained the classical Poiseuille
flow~\cite{Landau6}. It is well known that the Pouiseille flow is
stable against small perturbation for almost all Reynolds numbers;
The Poiseuille flow is laminar up to $Re\simeq 2\times 10^{4}$. This
flow is linearly stable for any $Re$, but is unstable for non-linear
perturbations at large $Re$: turbulent flow can be observed if,
e.g., one could force a large enough current. \footnote{A stability
analysis of the solution at large $Re$ numbers shows that the terms
of high order appear to be important, and one cannot consider only
linear perturbations (i.e., proportional to the field itself).
Assuming non-linear perturbations, one can then show that the
Pouiseille flow is unstable for an {\it arbitrary} (but small)
perturbation. This is different from the usual linear perturbation
case where the system is stable for {\it any} small linear
perturbation.} Since for an adiabatic constriction $1/k\sim 0$ (
i.e., at any given point the system is arbitrarily close to a pipe),
we conclude that {\it in an adiabatic QPC the flow is laminar for
almost any value of $Re$}. We note that compressibility of the
liquid may instead provide a lower critical $Re$ to observe
turbulence.

{\it Non-adiabatic QPCs -} Let us now look at a case where the
constriction is non-adiabatic.  A simple non-adiabatic potential for
which an analytical solution exists is $y=k|x|$ [see
Fig.~\ref{fig1}(b)]. Microscopically, the point $x,y=0$ could be,
e.g. a molecule sandwiched between two bulk electrodes and current
flows from one electrode to the other (see e.g.
\cite{DiVentra2002}). We are interested here in the dynamics close
to (but away from) this point.  The present problem could also be
solved by assuming a finite opening of the potential at the origin,
i.e. a potential of the form $y=k(|x|-\delta)$ ($\delta >0$). Our
conclusions would be unaffected by this finite opening. In the case
of the system in Fig.~\ref{fig1}(b) we know there is a critical $Re$
number, $R_{cr}$, determined by the angle $\tan (\beta/2)=1/k$,
above which the laminar flow is unstable. A simple calculation gives
the implicit relation between the angle $\beta$ and $R_{cr}$
\cite{Landau6}
\begin{eqnarray}
&&\beta(s)=\sqrt{1-2 s^2}K(s^2)\\
&&\frac{R_{cr}(s)}{6}=-\frac{1-s^2}{s^2}\beta(s)+\frac{\sqrt{1-2s^2}}{s^2}E(s^2)
\label{twoplanes}
\end{eqnarray}
where $K$ and $E$ are elliptic functions and $0<s<1$ is
an arbitrary parameter. This critical $Re$ is plotted in
Fig.~\ref{fig1} as a function of the angle $\beta$ and separates the
phase space in two regions: laminar for $Re<R_{cr}$, turbulent for
$Re>R_{cr}$. Equation (\ref{twoplanes}) has been again derived with the
no-slip condition at the boundary. A finite velocity at the boundary
(or even a larger velocity at the boundary than at the center of the
junction) would further reduce $R_{cr}$ for any $\beta$, i.e. it would
make the laminar solution even more unstable.  We can thus conclude that turbulence can
be observed in this case. Recently the current distribution for a 2DEG
QPC has been measured~\cite{Topinka2001}. An irregular
time-independent pattern has been observed and explained by the
presence of impurities in the system. We suggest that turbulent
effects could be observed in similar experiments on the 2DEG if
nonadiabaticity is introduced, e.g. by asymmetric electrodes that
would generate a potential of the type represented in
Fig.~\ref{fig1}(b).  In this case we can predict the dimension
$\lambda_{0}$ of the smallest observable eddies in the turbulent
regime. A simple dimensional analysis gives~\cite{Landau6}
$\lambda_{0}\sim l
\left({R_{cr}}/{R}\right)^{3/4}=l\left({I_{cr}}/{I}\right)^{3/4}$,
where $I_{cr}$ is the critical current, i.e., the current which, from
(\ref{twoplanes}) gives $R_{cr}$, and $l$ the linear size of the device. We easily see that
$\lambda_{0}$ decreases rapidly with increasing $\beta$.  The
viscosity $\eta$ can be evaluated through a perturbation theory on the
2D electron liquid~\cite{Conti1999}.  For a two dimensional electron
gas (2DEG) in a GaAs heterostructure ($m=0.067m_e$) with a sheet
density of $n\simeq 10^{15} ~\mathrm{m}^{-2}$ we have, $\eta/\hbar
n\simeq 0.05$ \cite{Conti1999}\footnote{Note that with these parameters the Fermi wavelength
$\lambda_{F}$ is about $80~\mathrm{nm}$.  Assuming a lateral dimension
$w=\lambda_{F}/2$ we get $\nu_c\simeq 10^{13}$ Hz, i.e. still orders
orders of magnitude faster than other inelastic scattering rates in
this system.}. For a typical current of
$1~\mathrm{\mu A}$, $Re=145$. We then expect that a turbulent flow is
developed for any angle larger than $\beta\sim \pi/10$. For instance,
for $\beta=\pi/2$ ($R_{cr}\sim 10$ from Fig.~\ref{fig1}), by assuming
a length $l$ of a typical device of about $1\mathrm{\mu m}$, we
evaluate the typical dimension of the smallest eddies to be
$\lambda_{0}\simeq 250~\mathrm{nm}$.  Our predictions should be thus
readily verifiable experimentally.

We finally conclude by noting that the geometry of nanoscale
structures, like, e.g. a molecule between bulk electrodes, is actually
closer to a conical structure for which the electron liquid is
turbulent for relatively small $Re$~\cite{Billiant1998}. The formation
of eddies in proximity to an atomic or molecular junction is thus much
more likely to occur than in the 2DEG case. Moreover, we expect that
the finite compressibility of the electron liquid will favour the
appearance of turbulent flow.

\ack
We acknowledge useful discussions with Na Sai, Neil
Bushong and Giovanni Vignale.  This work has been supported by the
Department of Energy grant DE-FG02-05ER46204.

\section*{References}
\bibliographystyle{vancouver}

\end{document}